\documentclass{PoS}
\newcommand{\Fermi}{\textit{Fermi}}

\title{Search for Gamma-ray Production in Supernovae Located in a Dense Circumstellar Medium with the \textit{Fermi}-LAT}

\ShortTitle{Gamma-rays from Supernovae}

\author{\speaker{Anna Franckowiak} on behalf of the \textit{Fermi}-LAT Collaboration\\
        Kavli Institute for Particle Astrophysics and Cosmology, SLAC National Accelerator Laboratory, Menlo Park, CA 94025, USA\\
        E-mail: \email{afrancko@slac.stanford.edu}}

\author{Kohta Murase\\
        Center for Particle and Gravitational Astrophysics; Department of Physics; Department of Astronomy \& Astrophysics, The Pennsylvania State University, University Park, Pennsylvania, 16802, USA\\
        E-mail: \email{murase@psu.edu}}
        
\author{Eran O.~Ofek on behalf of the PTF Collaboration\\
        Benoziyo Center for Astrophysics, Weizmann Institute of Science, 76100 Rehovot, Israel\\
        E-mail: \email{eran.ofek@weizmann.ac.il}}

\abstract{Supernovae (SNe) exploding in a dense circumstellar medium (CSM) are predicted to accelerate cosmic rays in collisionless shocks and emit GeV gamma rays and TeV neutrinos on a time scale of several months. Here we summarize the results of the first systematic search for gamma-ray emission in \textit{Fermi}-LAT data in the energy range from 100 MeV to 300 GeV from a large sample of SNe exploding in dense CSM. We search for a gamma-ray excess at the position of 147 SNe Type IIn in a one year time window after the optical peak time. In addition we combine the closest and optically brightest sources of our sample in a joint likelihood analysis in three different time windows (3, 6 and 12 months). No excess gamma-ray emission is found and limits on the gamma-ray luminosity and the ratio of gamma-ray to optical luminosity are presented.}

\FullConference{The 34th International Cosmic Ray Conference,\\
		30 July- 6 August, 2015\\
		The Hague, The Netherlands}

\begin{document}

\section{Introduction}

\begin{figure}
\begin{center}
\includegraphics[trim=0 5cm 3cm 0,clip, width=.7\textwidth]{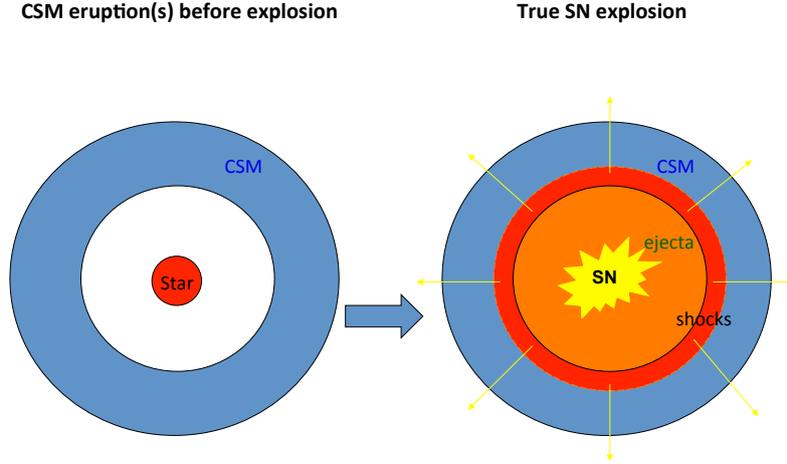}
\caption{Schematic picture of a SN exploding in dense circumstellar medium (taken from~\cite{Murase:2013kda}).}
\label{fig1}
\end{center}
\end{figure}

Supernovae (SNe) exploding in a dense circumstellar medium (CSM) are hypothesized to effectively accelerate cosmic rays (CR) in interactions of the SNe ejecta with the CSM. \cite{Murase:2010cu,Murase:2013kda} and~\cite{Katz:2011zx} showed that a collisionless shock is necessarily formed after the shock breakout if the SN progenitor is surrounded by an optically thick CSM. Cosmic rays might be accelerated to PeV energies and contribute to the knee structure in the cosmic-ray spectrum~\cite{Murase:2013kda,2003A&A...409..799S}. pp and p$\gamma$ interactions produce pions, which in the neutral case decay to gamma rays and in the charged case produce neutrinos in the decay chain. Interaction of the ejecta with the CSM can convert a large fraction of the kinetic energy to radiation causing this type of SNe to be more luminous than other core-collapse SNe~\cite{2002AJ....123..745R,2012ApJ...759..108S}. Figure~\ref{fig1} (taken from~\cite{Murase:2013kda}) shows a schematic picture of a SN exploding in dense circumstellar medium.

Several candidates for such interaction-powered SNe were found \cite{2007ApJ...659L..13O,2014ApJ...781...42O,2009AJ....137.3558S,Zhang2012}
and some superluminous supernovae were suggested to be powered by interactions \cite{Quimby:2009ps,Chevalier:2011ha}. Observationally, Type IIn SNe (``n'' for narrow) are the best candidates to be interacting with a dense CSM. They are spectrally characterized by the presence of strong narrow emission lines, mostly H$\alpha$, originating from surrounding H II regions. A slow spectral evolution indicates the presence of a dense CSM~\cite{1990MNRAS.244..269S,1997ARA&A..35..309F}.  They are often accompanied by precursor mass-ejection events, which could be responsible for the dense CSM~\cite{Ofek:2014ifa}. 

This conference proceedings summarize the work published in~\cite{2015arXiv150601647T}, where we perform the first systematic search for gamma-ray emission from Type IIn SNe in data recorded by the Large Area Telescope (LAT) on-board the \Fermi\,Gamma-ray Space Telescope in the energy range of $100\,$MeV to $300\,$GeV to probe interaction-powered SNe as cosmic-ray accelerators. We look for excess gamma-ray emission correlated in space and time with SNe found in optical surveys such as the Palomar Transient Factory (PTF)~\cite{Law:2009ys,Rau2009}. We use a sample of 147 Type IIn SNe. A full list of SNe used in this analysis can be found in~\cite{2015arXiv150601647T}.

To account for theoretical uncertainties on the expected $\gamma$-ray flux we perform a model independent search.

\section{Gamma-ray Data Analysis}

\begin{figure}
\begin{center}
\includegraphics[width=.45\textwidth]{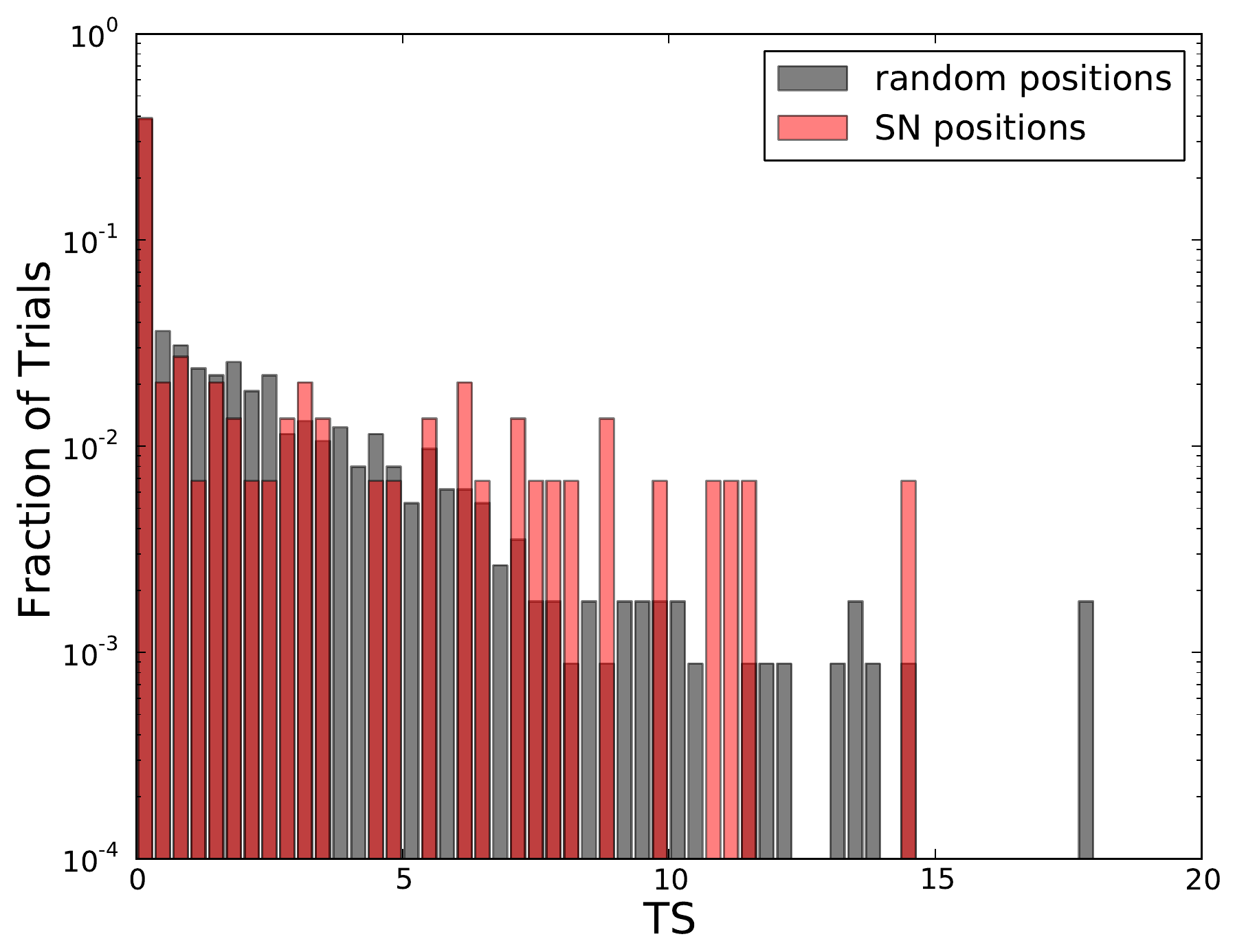}
\includegraphics[width=.45\textwidth]{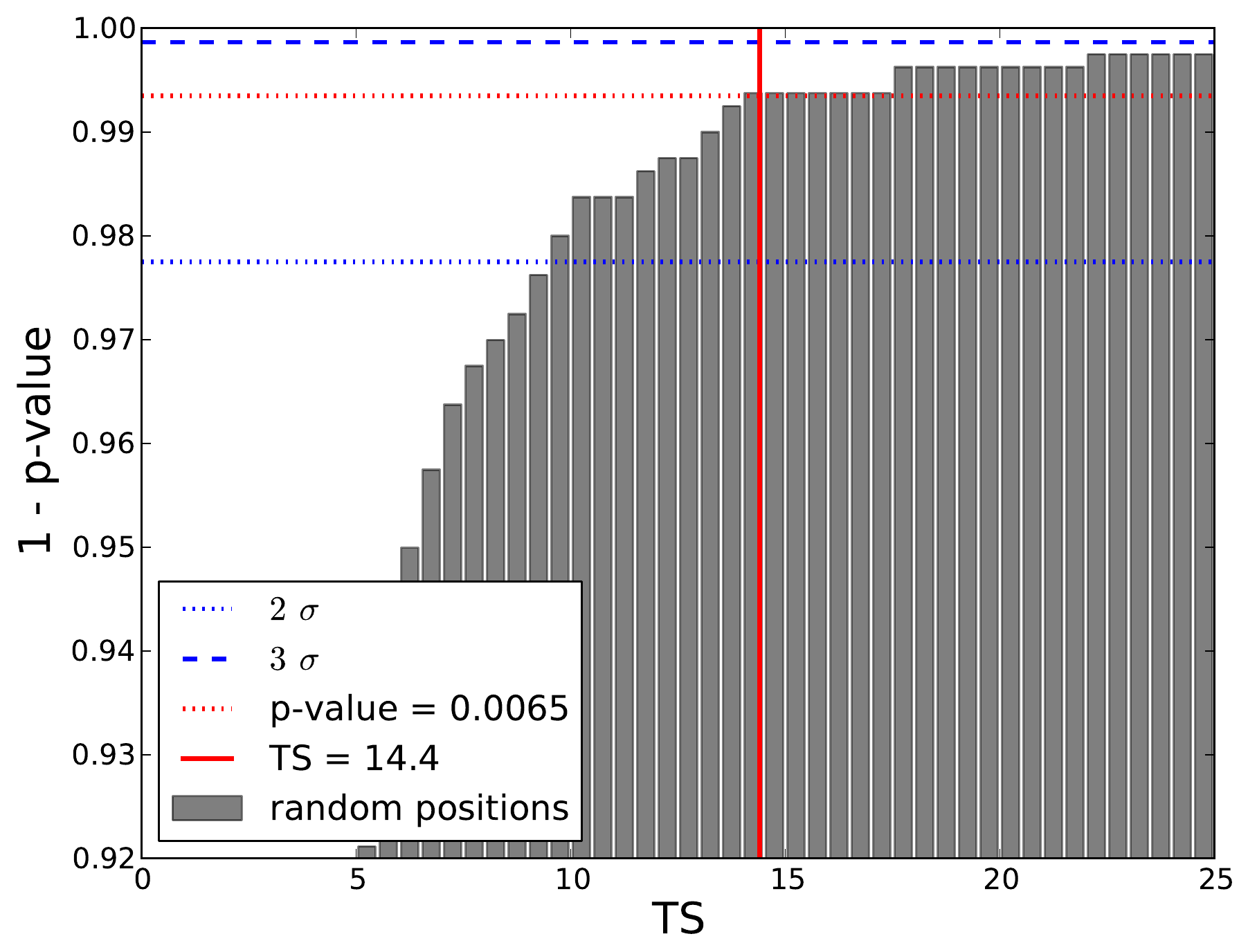}
\caption{Left: Distribution of SNe and random position TS values. Right: Cumulative distribution of random position TS values. The largest SN TS value is shown as vertical solid red line, it corresponds to a p-value of $0.0065$ and is compared to the $2\sigma$ and $3\sigma$ value. Considering trials factors, this p-value increases to $0.6$.}
\label{fig2}
\end{center}
\end{figure}

We analyze 57 month of \textit{Fermi}-LAT Pass 7  Reprocessed data. We select events of the source class and use the P7REP\_SOURCE\_V15 instrument response functions. First, we search for gamma rays from individual SNe, applying a binned likelihood analysis to a one year time window starting 4 weeks before optical peak time. We model the SN spectrum as a power-law leaving the normalization and index free. Our background model includes Galactic diffuse and isotropic gamma-ray emission in addition to point sources from the 2FGL source catalog~\cite{2012ApJS..199...31N}. For each SNe we calculate the test statistic (TS) defined as $TS = -2\Delta \log \mathcal{L}$, where $\Delta \log \mathcal{L}$ is the difference in likelihood between a model without including a source at the SN position (background-only model) and a model including a source with a power-law spectrum at the SN position. None of the SNe positions yields a significant TS value. In Fig.~\ref{fig2} we compare the SNe TS values with TS values obtained from repeating the analysis in random positions in the sky, which allows us to estimate the significance of each SN TS value. None exceeds the $3\sigma$ level. The largest found TS corresponds to a p-value of $0.0065$, which increases to $0.6$ considering trials factors.

\begin{figure}
\begin{center}
\includegraphics[width=.7\textwidth]{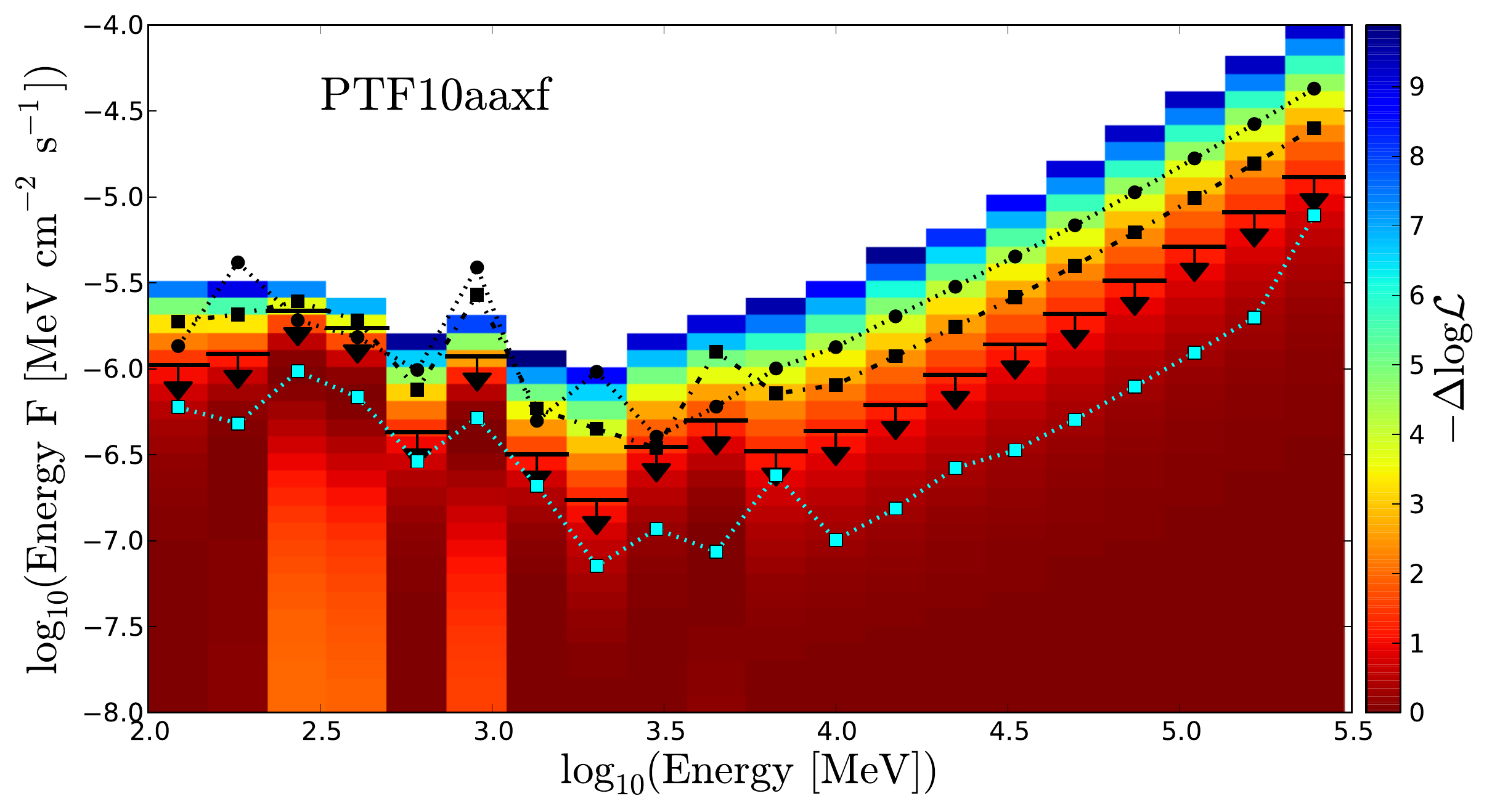}
\caption{Castro plot for SN\,2010jl: Within each bin, the color scale denotes the variation of the logarithm of the likelihood with respect to the best-fit value of the SN flux using a 12 months time window. 
$95\%$ CL upper limits are shown as black arrows for the 12 months time window and as dotted-dashed and dotted lines for a 6 and 3 months time window respectively. For the particular case of SN 2010jl we repeated the analysis for an extended time window spanning 4.5 years. The results are overlaid as cyan dotted line.}
\label{fig3}
\end{center}
\end{figure}

In a second step, we drop the assumption of a power-law spectral shape for the SNe and instead apply a bin-by-bin likelihood analysis, where we allow the normalization of the SN flux to float in each energy bin while the background parameter (i.e., background point sources, diffuse and isotropic emission) are fixed to values obtained from the broad-band fit. We repeat this analysis for three different time windows: 3, 6 and 12\,months. The results for one source, SN\,2010jl, are displayed in form of the ``Castro plot'' in Fig.~\ref{fig3} (similar plots for the other SNe can be found in~\cite{2015arXiv150601647T}). The Castro plot shows the bin-by-bin likelihood function used to test for a putative gamma-ray source at the position of the SN. It allows us to fit any kind of spectral model to the results without repeating the gamma-ray data analysis, i.e. the results are independent of any kind of spectral model assumption. The bin-by-bin likelihood results can be used to re-create a global likelihood for a given signal spectrum by tying the signal parameters across the energy bins (see~\cite{2015arXiv150601647T} and \cite{Ackermann:2013yva} for details).

\begin{figure}
\includegraphics[width=.5\textwidth]{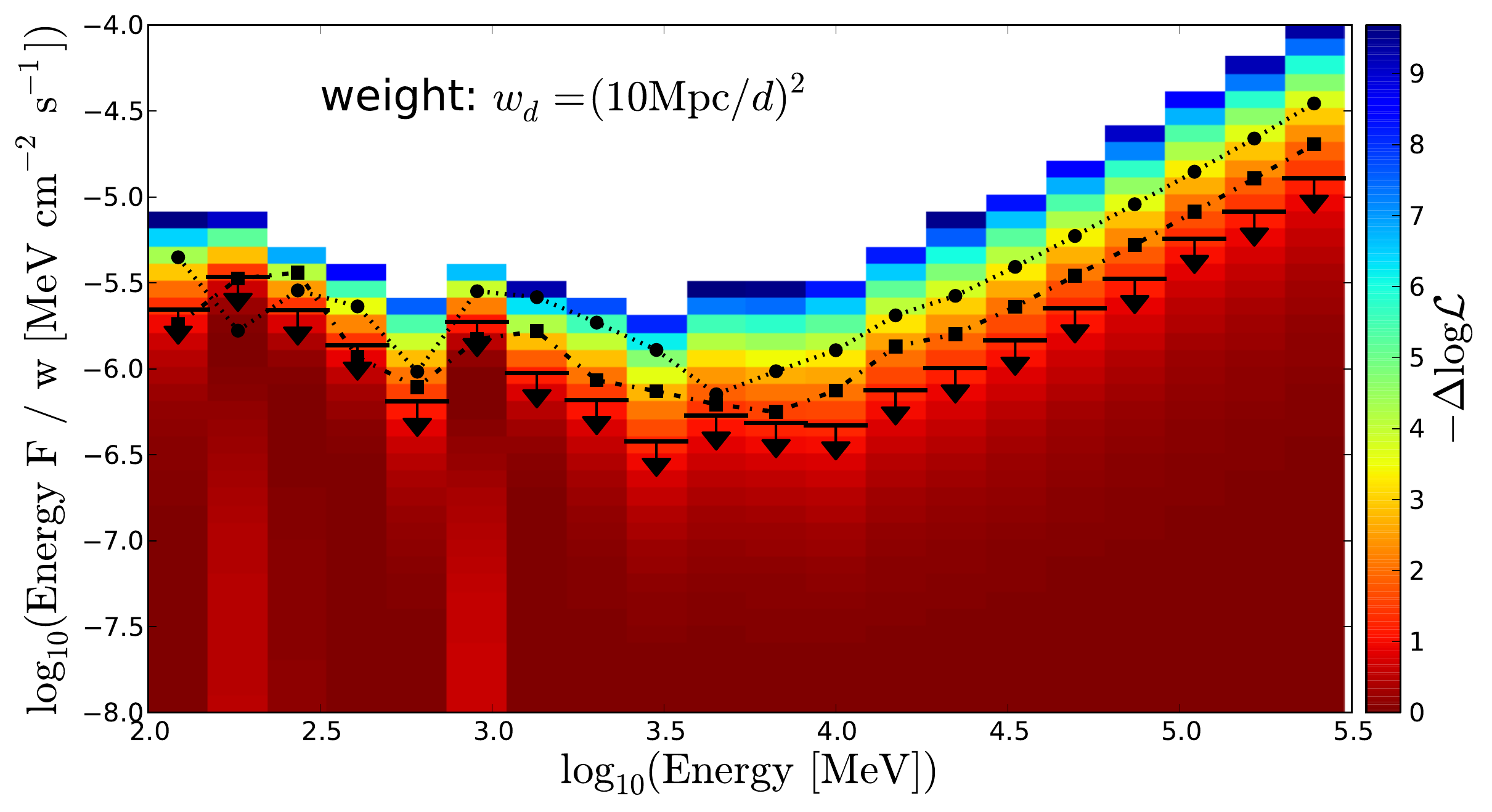}
\includegraphics[width=.5\textwidth]{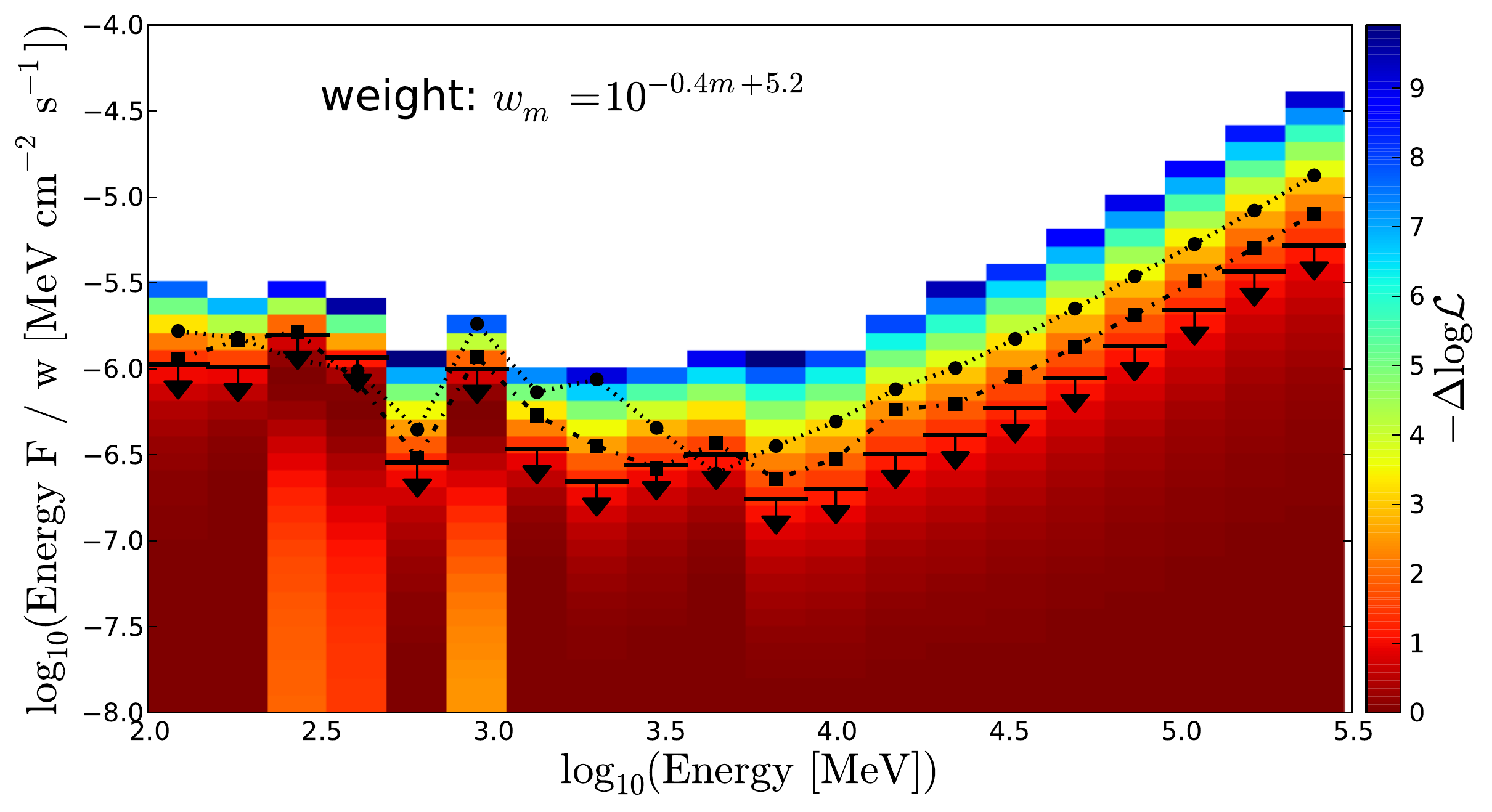}
\caption{Castro plot for joint likelihood results. Left using a $1/d^2$ weighting. Right using a $10^{-0.4m}$ weighting. Within each bin, the color scale denotes the variation of the logarithm of the likelihood with respect to the best-fit value of the SN flux using a 12 months time window. $95\%$ CL upper limits are shown as black arrows for the 12 months time window and as dotted-dashed and dotted lines for a 6 and 3 months time window respectively.}
\label{fig4}
\end{figure}

To be more sensitive to a weak gamma-ray signal we combine the sources in a joint likelihood analysis. For this analysis we select the 16 optically brightest and closest sources, fulfilling $z<0.015$ or $m<16.5$, where $z$ is the redshift and $m$ the optical peak magnitude\footnote{For some SNe the optical peak magnitude was not determined. In those cases we use the discovery time and magnitude.}. We repeat the joint likelihood analysis for 3 different time windows (3, 6 and 12 months) and for two different weighting schemes. In one case we weight each source with $1/d^2$, i.e., we assume that all sources have a similar gamma-ray luminosity. In the other case we weight with a weight proportional to $10^{-0.4m}$, i.e., assuming that the gamma-ray luminosity is correlated with the optical luminosity. We produce the Castro plot for both cases and fit the results with a power-law and power-law with exponential cut-off spectral model (see Fig.~\ref{fig4}). None significantly improves the likelihood of the fit.

\section{Results}

We use our flux limits presented above to set limits on the gamma-ray luminosity and the ratio of gamma-ray to optical luminosity (for details of the limit calculation see~\cite{2015arXiv150601647T}). We use a gamma-ray emission model based on~\cite{Murase:2010cu}, which assumes that gamma-ray emission is produced by CRs accelerated at the early collisionless shock between SN ejecta and the dense CSM. The gamma-ray emission can be predicted following the procedure outlined in \cite{Murase:2013kda}, if the model parameters are determined by optical and X-ray observations. However, such model-dependent analyses was deferred to future work. Instead in \cite{2015arXiv150601647T} we take a model-independent approach aiming to constrain the gamma-ray luminosity as a function of the proton spectral index $\Gamma$. We assume that the CR proton spectrum follows a power-law function in momentum and in the calorimetric limit the gamma-ray spectral index follows the proton spectral index. We do not take into account the effect of gamma-ray absorption, but ~\cite{Murase:2013kda} showed that GeV gamma-rays can escape without severe matter attenuation if the shock velocity is high enough.

As an example we show the predicted gamma-ray flux for the optically brightest source of the sample, SN\,2010jl using a generic gamma-ray flux model. A more detailed modeling of this source based on multi-wavelength observations will follow in future work. Here we assumed a proton spectral index of $\Gamma = -2$ and a normalization of the gamma-ray flux that yields to a ratio of gamma-ray to optical luminosity between 0.01 and 0.1.
Furthermore we assumed a distance of $48.7$\,Mpc and an apparent R-band peak magnitude of $13.2$. Our resulting flux upper limit touches the optimistic model prediction, i.e. brings us into reach of the interesting parameter space. Better constraints on the gamma-ray escape fraction are crucial to calculate stringent limits on the proton acceleration efficiency and will be obtained in more detailed modeling.

\begin{figure}
\begin{center}
\includegraphics[width=.6\textwidth]{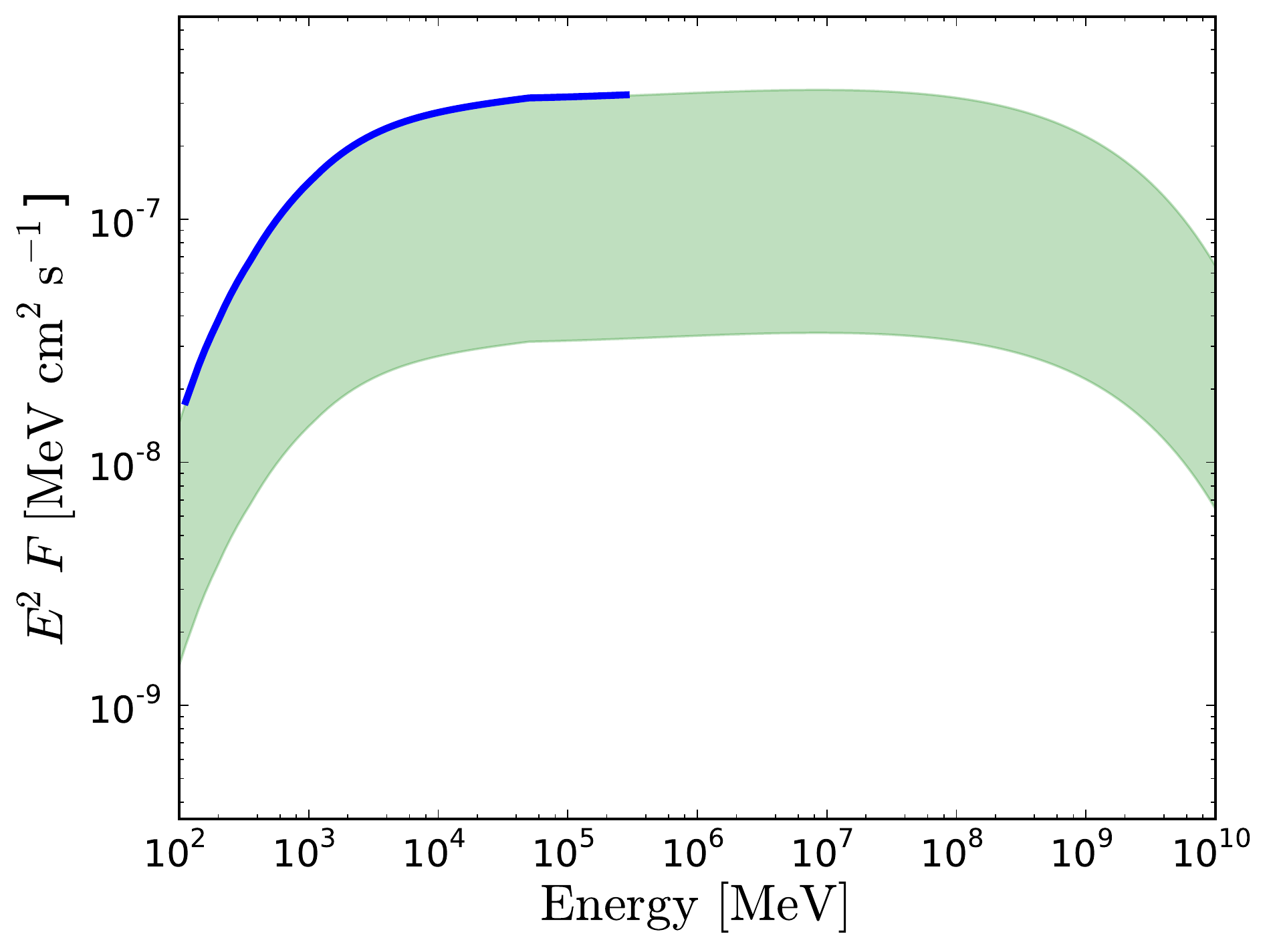}
\caption{Predicted gamma-ray flux using a generic gamma-ray emission model for SN\,2010jl including theoretical uncertainties (shown as green band) compared to $95\%$ CL flux upper limits obtained from analyzing a one year time window in blue.}
\label{fig5}
\end{center}
\end{figure}

\begin{figure}
\includegraphics[width=.5\textwidth]{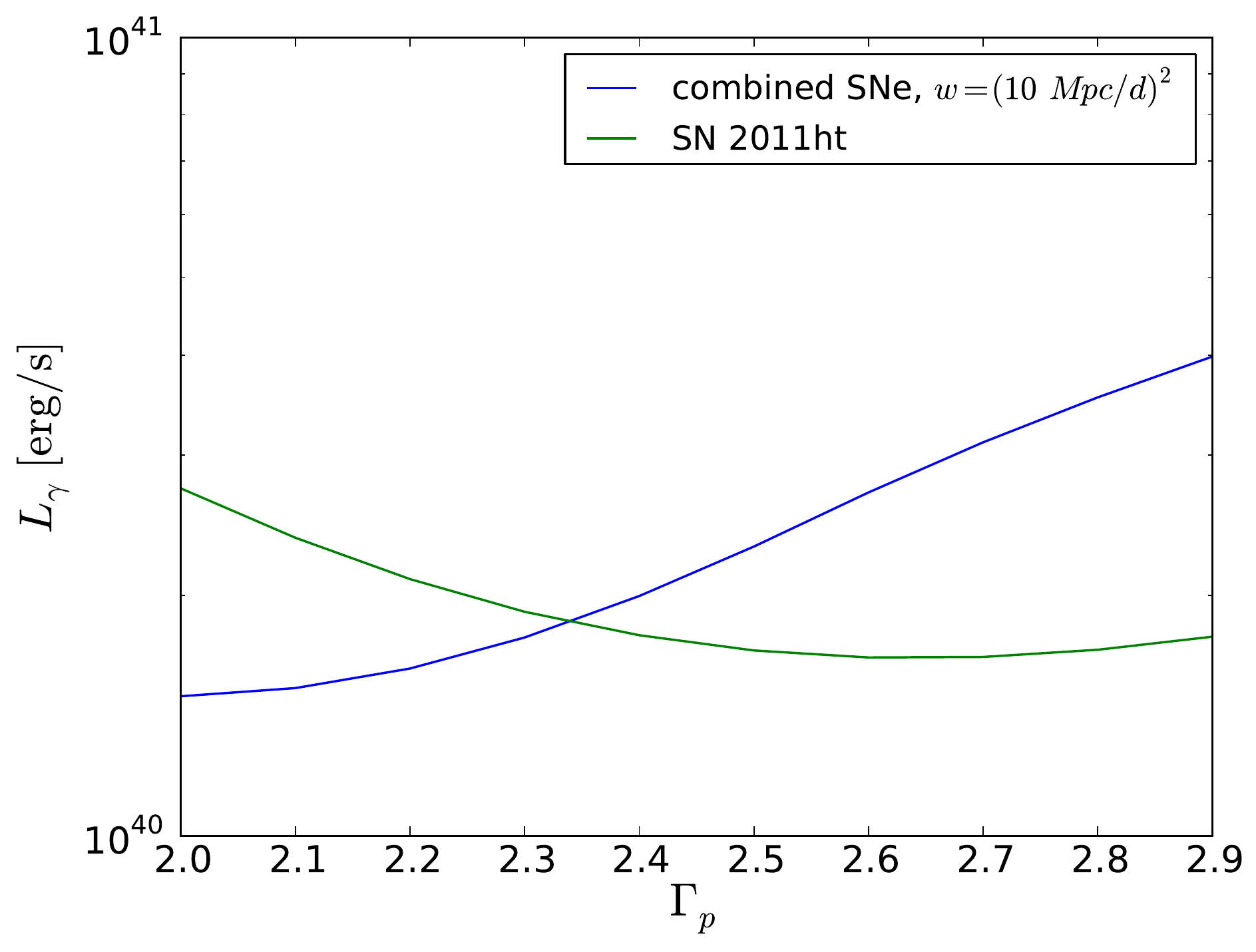}
\includegraphics[width=.5\textwidth]{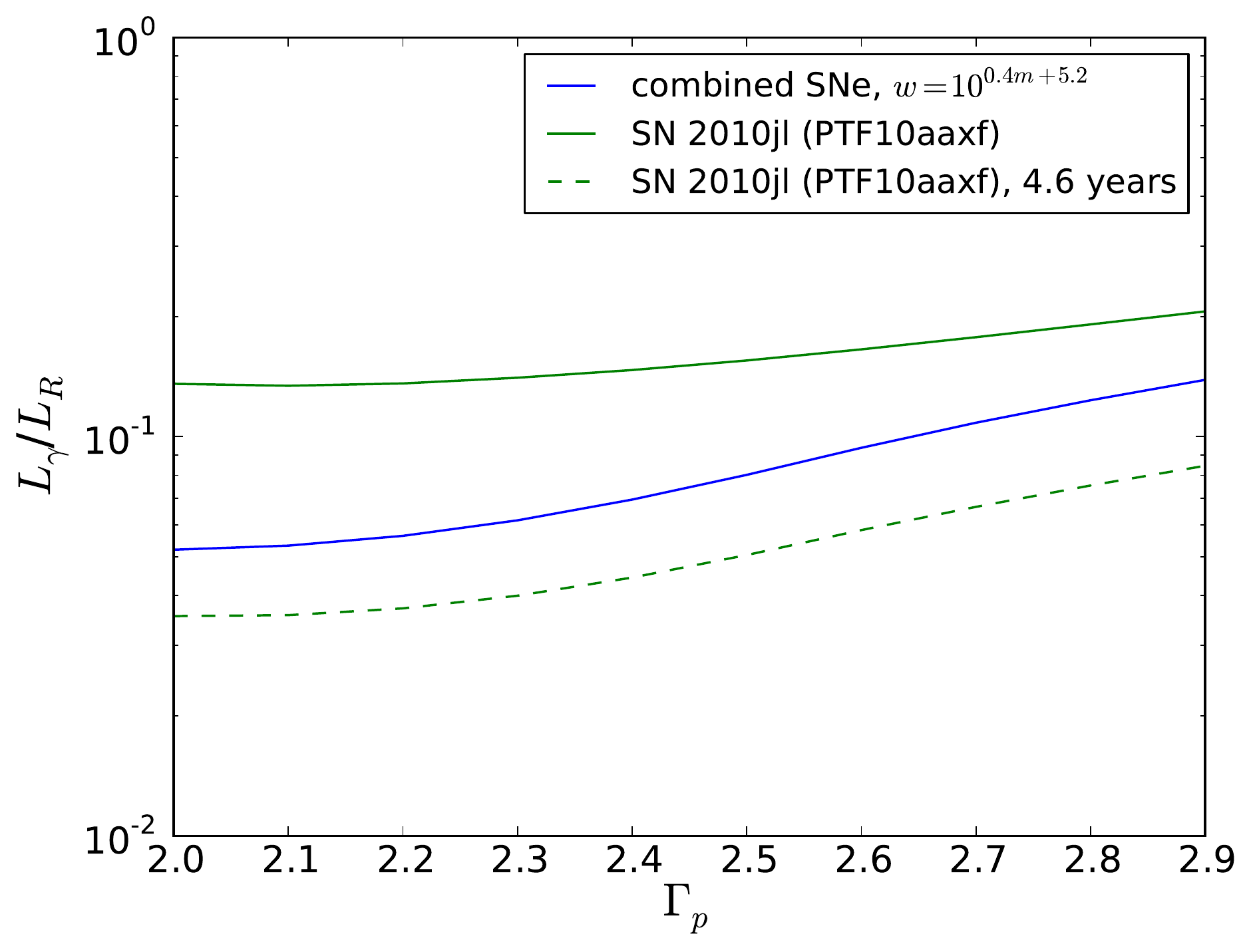}
\caption{Left: Limits on the gamma-ray luminosity using the joint likelihood results assuming that all sources have the same gamma-ray luminosity compared to the limit obtained from analyzing only the closest source in our sample (SN\,2011ht). Right: Limits on the ratio of gamma-ray to optical luminosity based on the joint likelihood results assuming a correlation of optical to gamma-ray luminosity compared to the single-source limit from the optically brightest source of the sample (SN\,2010jl).}
\label{fig6}
\end{figure}

In Fig.~\ref{fig6} (left panel) we show our limit on the gamma-ray luminosity assuming that all sources have the same gamma-ray luminosity $L_\gamma$ compared to the limit obtained from the analysis of the closest SN in our sample (SN\,2011ht with a distance of $d = 17.7$\,Mpc). For some choices of $\Gamma$ the single source limit is better than the combined limit, indicating a statistical under-fluctuation in
the individual analysis of this source or an over-fluctuation in one of the sources included in the joint likelihood. We also show the limit on the ratio of gamma-ray to optical luminosity $L_\gamma / L_R$, where $L_R$ is the R-band luminosity (see Fig.~\ref{fig6}, right panel). Note that $L_\gamma / L_R$ is an upper bound on $L_\gamma / L_{\textrm{rad}}$, where $L_{\textrm{rad}}$ is the bolometric radiation luminosity. We compare the joint likelihood limit to the limit obtained using only the optically brightest source of our sample, SN\,2010jl.

\section{Conclusion}

The origin of the multi-wavelength emission of Type IIn SNe and the onset of CR production in SN remnants is not fully understood. Type IIn SNe are expected to be host
sites of particle acceleration, which could be pinpointed by transient gamma-ray signals. In this conference proceedings we presented the results of~\cite{2015arXiv150601647T}, where we for the first time apply a systematic search for gamma-ray emission from a large ensemble of Type IIn SNe in coincidence with optical signals. We did not find evidence for a signal, but our limits start to reach the expected model parameter ranges. 

We use a joint likelihood analysis to set stringent limits on the gamma-ray luminosity and the ratio of gamma-ray to optical luminosity. For proton spectral indices of $\Gamma < 2.7$ we can exclude at $95\%$ CL that $L_\gamma / L_R > 0.1$ assuming that $L_\gamma / L_R$ is constant. Future model-dependent calculations based on multi-wavelength observations will allow us to also set stringent limits on the proton acceleration efficiency. 

The limits presented here are based on minimal assumptions about the gamma-ray production and can be used to test various models.

\acknowledgments
The \textit{Fermi}-LAT Collaboration acknowledges generous ongoing support
from a number of agencies and institutes that have supported both the
development and the operation of the LAT as well as scientific data analysis.
These include the National Aeronautics and Space Administration and the
Department of Energy in the United States, the Commissariat \`a l'Energie Atomique
and the Centre National de la Recherche Scientifique / Institut National de Physique
Nucl\'eaire et de Physique des Particules in France, the Agenzia Spaziale Italiana
and the Istituto Nazionale di Fisica Nucleare in Italy, the Ministry of Education,
Culture, Sports, Science and Technology (MEXT), High Energy Accelerator Research
Organization (KEK) and Japan Aerospace Exploration Agency (JAXA) in Japan, and
the K.~A.~Wallenberg Foundation, the Swedish Research Council and the
Swedish National Space Board in Sweden.

Additional support for science analysis during the operations phase is gratefully
acknowledged from the Istituto Nazionale di Astrofisica in Italy and the Centre National d'\'Etudes Spatiales in France.

This paper is based on observations obtained with the Samuel Oschin Telescope as part of the Palomar Transient Factory project, a scientific collaboration between the California Institute of Technology, Columbia University, Las Cumbres Observatory, the Lawrence Berkeley National Laboratory, the National Energy Research Scientific Computing Center,                                                                                                                                            
the University of Oxford, and the Weizmann Institute of Science. Some of the data presented herein were obtained at the W. M. Keck Observatory, which is operated as a scientific partnership among the California Institute of Technology, the University of California, and NASA; the Observatory was made possible by the generous financial support of the W. M. Keck Foundation.  We are grateful for excellent staff assistance at Palomar, Lick, and Keck Observatories.  E.O.O. is incumbent of the Arye Dissentshik career development chair and is grateful to support by grants from the Willner Family Leadership Institute Ilan Gluzman (Secaucus NJ), Israeli Ministry of Science, Israel Science Foundation, Minerva and the I-CORE Program of the Planning and Budgeting Committee and The Israel Science Foundation.     

\bibliography{SN_paper}

\end{document}